\documentclass[twocolumn,pra,aps,superscriptaddress]{revtex4}
\usepackage[T1]{fontenc}
\usepackage[latin9]{inputenc}
\usepackage{amsmath}
\usepackage{amssymb}
\usepackage{makecell}
\usepackage{graphicx}
\usepackage{booktabs}
\usepackage{subfigure}
\usepackage{esint}

\newcommand{\Rmnum}[1]{\expandafter\@slowromancap\romannumeral #1@}

\makeatletter
\@ifundefined{textcolor}{}
{%
 \definecolor{BLACK}{gray}{0}
 \definecolor{WHITE}{gray}{1}
 \definecolor{RED}{rgb}{1,0,0}
 \definecolor{GREEN}{rgb}{0,1,0}
 \definecolor{BLUE}{rgb}{0,0,1}
 \definecolor{CYAN}{cmyk}{1,0,0,0}
 \definecolor{MAGENTA}{cmyk}{0,1,0,0}
 \definecolor{YELLOW}{cmyk}{0,0,1,0}
}

\makeatother

\begin{document}
\hyphenation{PPT} \hyphenation{start} \hyphenation{cavi-ties} \hyphenation{func-tions}

\title{Enhancing the sensitivity of rotation in a multi-atom Sagnac interferometer}

\author{Fei Yao$^1$, Yanming Che$^1$, Yuguo Su$^1$, Hongbin Liang$^1$, Jiancheng Pei$^1$ and Xiaoguang Wang}

\email{xgwang1208@zju.edu.cn}

\affiliation{Zhejiang Institute of Modern Physics, Department of Physics, Zhejiang
University, Hangzhou 310027, China}
\begin{abstract}

We investigate quantum sensing of rotation with a multi-atom Sagnac interferometer and present multi-partite entangled states to enhance the sensitivity of rotation frequency.
For studying the sensitivity, we first present a Hermitian generator with respect to the rotation frequency. The generator, which contains the Sagnac phase, is a linear superposition of a $z$ component of the collective spin  and a quadrature operator of collective bosons depicting the trapping  modes, which enables us to conveniently study the quantum Fisher information (QFI) for any initial states.
With the generator, we derive the general QFI which can be of square dependence on the particle number, leading to Heisenberg limit.
And we further find that the QFI may be of biquadratic dependence on the radius of the ring which confines atoms, indicating that larger QFI is achieved by enlarging the radius.
In order to obtain the square and biquadratic dependence, we propose to use partially and globally entangled states as inputs to enhance the sensitivity of rotation.

\end{abstract}
\maketitle
\section{Introduction}\label{sec:intro}
The Sagnac phase, first discussed by Sagnac in 1913~\cite{Sagnac1913}, is the phase difference between two counter-propagating waves around a closed-loop in a rotating frame~\cite{Chow1985}. In general, the Sagnac interference can be realized via  fibre-optic gyroscopes~\cite{Arditty1981} and atomic gyroscopes~\cite{Cronin2009}, which measure the rotation rate relative to a inertial reference.
In recent years, improving the performance of atom interferometers has attracted lots of attention~\cite{Gustavson2000,Durfee2006,Dickerson2013}, because atom interferometers are more sensitive than optical Sagnac gyroscope in the measurements of the physical quantities. Meanwhile, atom interferometers have a wide range of applications in the test of weak equivalence
principle~\cite{Zhou2015,Geiger2018,Overstreet2018}, the measurement of gravity~\cite{Peters1999,Hu2013,Louchet2011,Wang2017,Sorrentino2014}, the inertial navigation~\cite{Stockton2011,Berg2015,Lenef1997,Geiger2011,Dutta2016} and the measurement of fundamental physical constants~\cite{Lamporesi2008,Chiow2009}.

In atom gyroscopes, the Sagnac effect plays an important role, which enables the rotation measurements with high precision. The study on rotation sensing is essential in this field. Recently, a scheme for Sagnac interferometry with a single atomic clock was proposed in Ref.~\cite{Stevenson2015}, where atoms are guided by two potential wells moving around a closed loop in opposite directions together with the use of Ramsey sequences, and the Sagnac phase was finally read out by the measuring the particle number difference. Based on this work,  the accuracy of rotation sensing in the Sagnac interferometer with multi-particle states was studied in Ref.~\cite{Luo2017}. The authors investigated the rotation measurement precision in the multi-particle atom interferometer via calculating the QFI.

Inspired by these works, we consider how to further enhance the sensitivity of rotation in a multi-atom Sagnac interferometer. In order to improve the sensitivity of rotation, we estimate the precision of rotation frequency with the help of the QFI.
As we know, for a single parameter, the Cram\'er-Rao inequality provides us a lower bound on the variance of an unbiased estimator~\cite{Helstrom1976,Holevo1982}. In quantum metrology, the precision of an parameter$\phi$ is determined by the quantum Cram'er-Rao bound (QCRB)
\begin{equation}\label{eq:QCRB}
\delta\phi\geq\frac{1}{\sqrt{F(\phi)}},
\end{equation}
 where $F$ is the QFI~\cite{Braunstein1994,Braunstein1996,Luo2003,Pezz2009}.

For a unitary parametrization transformation $U=\exp(-\mathrm{i}tH(\phi))$, the parametrized state can be expressed as $\rho(\phi)=U(\phi)\rho_{0}U^\dagger(\phi)$, where $\rho_{0}$ is an initial state independent of $\phi$.
When $\rho_0=|\psi\rangle\langle\psi|$ is a pure state, the QFI with respect to $\phi$  is given by~\cite{Liu2015,Liu2014,Liu2013}
\begin{equation}\label{eq:QFI}
F=4(\left\langle\psi\right|{\cal{H}}^2_\phi\left|\psi\right\rangle-{\left\langle\psi\right|{\cal{H}}_\phi\left|\psi\right\rangle}^2),
\end{equation}
where
\begin{equation}\label{eq:generator}
{\cal{H}}_\phi= \mathrm{i}(\partial_{\phi}U^{\dagger})U
\end{equation}
is a Hermitian generator with respect to $\phi$, which is independent of initial state~\cite{Liu2015}. In order to study the QFI for any initial states, it is convenient to first obtain the generator ${\cal{H}}_{\phi}$.

In this paper, we derive a Hermitian generator ${\cal{H}}_{M}$ with respect to the rotation frequency in multi-atom Sagnac interferometer.
With generator ${\cal{H}}_{M}$, we give the general expression of QFI for any pure states in terms of correlation functions. It is found that the general QFI is a linear superposition of particle number $N$ and the square $N^2$.  In order to improve the rotation sensitivity, we attempt to search  appropriate initial states which gives a larger coefficient before $N^2$.

In Ref.~\cite{Simon2016}, Haine evaluated the sensitivity in matter wave interferometer with the classical and quantum Fisher information. For high spatial resolution, the author used both the spin and spatial degrees of freedom. And a general multi-particle state was introduced,
\begin{equation}
|\psi\rangle=\bigotimes_{k=1}^N\left(\left|\uparrow\right\rangle_k \otimes \left|\psi_{\uparrow}\left(\textbf{r}\right)\right\rangle_k+\left|\downarrow\right\rangle_k \otimes \left|\psi_{\downarrow}\left(\textbf{r}\right)\right\rangle_k\right)
\end{equation}
where $\left|\uparrow\right\rangle_k$ and $\left|\downarrow\right\rangle_k$ are two spin states for $k$-th spin, $\left|\psi_{\uparrow}\left(\textbf{r}\right)\right\rangle$ and $\left|\psi_{\downarrow}\left(\textbf{r}\right)\right\rangle$ are states manipulated independently in two trapping potentials. Obviously, this state is only locally entangled.
For achieving the Heisenberg limit, inspired by this work, we propose to use the following type of the multi-particle globally entangled state
\begin{equation}\label{psi1}
|\psi\rangle=\frac{1}{\sqrt{2}}\left(\bigotimes_{k=1}^N\left|\uparrow,\psi_{\uparrow}\right\rangle_k+
\bigotimes_{k=1}^N\left|\downarrow,\psi_{\downarrow}\right\rangle_k\right).
\end{equation}
This state displays spin-spin, space-space, and spin-space entanglement. We will see that this globally entangled state has advantages over others in enhancing the rotation sensitivity.

This paper is organized as follows. In Sec.~\uppercase\expandafter{\romannumeral2}, we derive
the generator ${\cal{H}}_{M}$ and the general expression of QFI with respect to the rotation frequency.
The QFI is expressed in terms of various correlations and explicit dependence on the total particle number is given.
In Sec.~\uppercase\expandafter{\romannumeral3}, with the general QFI, we calculate and compare the QFI for partially entangled state and globally entangled state.
A summary is given in Sec.~\uppercase\expandafter{\romannumeral4}.
\section{The general QFI with respect to rotation frequency}
In this section, we first derive the generator with respect to rotation frequency in multi-atom Sagnac interferometer. And using this generator, we give a general QFI for any initial pure states, which is a linear superposition of particle number $N$ and $N^2$.
\subsection{The generator with respect to rotation frequency}
From Ref.~\cite{Stevenson2015}, the Hamiltonian of the atom interferometer with a single particle is given by
\begin{equation}\label{eq:H2}
H\left(t\right)=\hbar\omega a^{\dagger} a+\mathrm{i}p_{c} r\left[\Omega+{\sigma}_{z}\omega_{p}(t)\right]\left(a-{a}^{\dagger}\right),
\end{equation}
where $\omega$ is the trapping frequency of two harmonic potentials. $a^{\dagger} \left(a\right)$ is the creation (annihilation) operator for the trap mode. For the second term in Eq.~\eqref{eq:H2}, $p_{c}=\sqrt{\frac{m\hbar\omega}{2}}$ is the characteristic  momentum of system, $r$ is the radius of the ring which confines the atoms, $\Omega$ is the angular frequency of laboratory frame, and $\omega_{p}(t)$ is the angular speed of the two harmonic potentials that rotating in opposite directions.
Operator $\sigma_{z}$ is pseudo-spin operator satisfying $\sigma_{z}\left|\uparrow\right\rangle\left(\left|\downarrow\right\rangle\right)=\left|\uparrow\right\rangle\left(-\left|\downarrow\right\rangle\right)$.

For the unitary operator generated by the above Hamiltonian, the corresponding generator of the QFI with respect to $\Omega$ in the single-atom Sagnac interferometer is obtained as (see Appendix A)
\begin{equation}\label{eq:Hs}
\mathcal{H}=T_{C}\left[C_1(\tau)a^{\dagger}+h.c.\right]+\frac{1}{\omega}C_{0}(\tau)+T_{S}C_2(\tau)\sigma_{z},\\
\end{equation}
where
\begin{eqnarray}
T_{C}&=&r\sqrt{\frac{2m}{\omega\hbar}},\label{Tc}\\
T_{S}&=&\frac{\phi_{s}}{\Omega}=\frac{2m\pi r^2 }{\hbar} \label{Ts},\\
C_{0}(\tau)&=&\frac{\phi_{s}}{2\pi}\left(\omega \tau-\sin\omega\tau\right),\\
C_1(\tau)&=&\mathrm{i}\sin{\left(\frac{\omega \tau}{2}\right)}\exp\left(\frac{\mathrm{i}\omega \tau}{2}\right) ,\label{C1}\\
C_2(\tau)&=&\frac{1}{2}\left(1-\frac{1}{\pi}\int_{0}^{\tau}\omega_{p}\left(t\right)\cos\left[\omega(t-\tau)\right]dt\right) \label{C2}.\\
\nonumber
\end{eqnarray}
Here, $T_{C}$ is the intrinsic time of the system. $T_{S}={\phi_{s}}/{\Omega}$ can be viewed as the Sagnac time, where $\phi_{s}={2m\Omega \pi r^2 }/{\hbar}$ is the well-known Sagnac phase \cite{Stevenson2015,Simon2016,Che}. And $\tau$ is the total evolution time, which satisfies~\cite{Stevenson2015}
\begin{equation}\label{eq:wp}
\int_{0}^{\tau}{\omega_{p}(t)}dt=\pi.
\end{equation}
Using Eq.~\eqref{eq:wp}, one can obatin $C_2(\tau)\geq0$.
Based on Eq.~\eqref{eq:QFI}, we can obtain the corresponding QFI with respect to $\Omega$ in single-atom Sagnac interferometer.

Above we derived the generator ${\cal{H}}$ in the single-atom Sagnac interferometer.
Next, we will study the generator of QFI with respect to $\Omega$ in multi-atom Sagnac interferometer. For multi-atom Sagnac interferometer, the Hamiltonian is given by~\cite{Luo2017}
\begin{equation}\label{eq:HM}
H_{M}\left(t\right)=\sum_{k=1}^{N}H^{(k)}\left(t\right),\\
\end{equation}
\begin{small}
where
\begin{equation}
H^{(k)}\left(t\right)=\hbar\omega a^{\dagger}_{k} a_{k}+\mathrm{i}p_{c}r\left[\Omega+{\sigma}_{z}^{\left(k\right)}\omega_{p}(t)\right]\left(a_{k}-{a}^{\dagger}_{k}\right)\nonumber\\
\end{equation}
\end{small}
is the Hamiltonian for $k$-th particle.

For this Hamiltonian, the corresponding generator ${\cal H}_{M}$ of the QFI with respect to $\Omega$ can be given by (see Appendix A)
\begin{equation}\label{eq:calHM}
{\cal H}_{M}= {\cal H}_{1}+{\cal{H}}_{0}+{\cal H}_{2},\\
\end{equation}
where
\begin{eqnarray}
{\cal H}_{1}&=& T_{c}\sum^{N}_{k=1}X_{k},\label{calH1}\\
{\cal{H}}_{0}&=&\frac{N}{\omega}C_{0}(\tau),\label{calH0}\\
{\cal H}_{2}&=&T_{s}C_2(\tau)J_{z}\label{calH2}.
\end{eqnarray}
Here, $J_{z}=\sum^{N}_{k=1}\sigma_{z}^{\left(k\right)}$ is $z$ component of collective spin operator.
$X_{k}=\left[ C_1(\tau)a^{\dagger}_{k}+h.c.\right]$ is a quadrature of the $k$-th bosonic mode.
Thus, ${\cal H}_{1}$ is just the collective quadrature operator for $N$ trapping bosonic modes.

\subsection{QFI in terms of Correlation functions}
In above section, we have obtained generator ${\cal{H}}_{M}$.
Based on this generator and Eq.~\eqref{eq:QFI}, one can get the QFI with respect to $\Omega$ for any initial pure states in the multi-particle scheme (see Appendix B),

\begin{eqnarray}
F&=&4\left[\mathrm{Var}({\cal{H}}_{1})+\mathrm{Var}({\cal{H}}_{2})+2\mathrm{Cov}({\cal{H}}_{1}{\cal{H}}_{2})\right]\label{FM}\\
 & = & 4\left[(\beta-\gamma)N+\gamma N^2\right],\label{FMM}
\end{eqnarray}
where
\begin{eqnarray*}
\beta &=& T_{C}^{2}\mathrm{Var}(X_{1})+T_{S}^2 C_{2}^2(\tau)\mathrm{Var}(\sigma_{z}^{(1)}) \\
&&+2T_{C}T_{S}C_{2}(\tau)\mathrm{Cov}(X_{1},\sigma_{z}^{(1)}),  \\
\gamma &=& T_{C}^2\mathrm{Cov}(X_{1},X_{2})+T_{S}^2 C_{2}^2(\tau)\mathrm{Cov}(\sigma_{z}^{(1)},\sigma_{z}^{(2)})\\
&&+2T_{C}T_{S}C_{2}(\tau)\mathrm{Cov}(X_{1},\sigma_{z}^{(2)}).
\end{eqnarray*}
From Eq.~\eqref{FM},  one can see that as ${\cal{H}}_{0}$ is a $c$-number, it has no contribution to the QFI. In other words, the QFI only depends on ${\cal{H}}_{1}$ and ${\cal{H}}_{2}$, which are the bosonic and spin operators, respectively. In Eq.~\eqref{FMM}, because $T_{C}$, $T_{S}$, and $C_{2}(\tau)$ are all non-negative numbers, if one of the correlation functions in the expression of $\gamma$ are positive, the QFI
will depends on $N^2$. It implies that the ultimate measurement precision can reach the Heisenberg limit, due to the QCRB. Therefore, one can adopt appropriate states to guarantee that all the correlation functions in $\gamma$ are positive to get larger QFI.

In addition, from Eqs.\eqref{Tc}-\eqref{C2}, the QFI in Eq.~\eqref{FMM} can be rewritten as a polynomial of $R=r/{\rho}$, with ${\rho}=\sqrt{\hbar/m \omega}$ being the
harmonic oscillator length,
\begin{equation}\label{FMr}
F=\lambda_{1}R^2+\lambda_2 R^3+\lambda_3 R^4,
\end{equation}
where
\begin{small}
\begin{eqnarray}
\lambda_{1}&=&\frac{2}{\omega^2}N\left[\mathrm{Var}(X_{1})+(N-1)\mathrm{Cov}(X_{1},X_{2})\right],\nonumber\\
\lambda_{2}&=&\frac{\sqrt{2}\pi}{\omega^2} C_{2}(\tau)N\big[\mathrm{Cov}(X_{1},\sigma_{z}^{(1)})+\left(N-1\right)\mathrm{Cov}(X_{1},\sigma_{z}^{(2)})\big],\nonumber\\
\lambda_{3}&=&\frac{4\pi^2}{\omega^2}C_{2}^2(\tau)N\left[\mathrm{Var}(\sigma_{z}^{(1)})+(N-1)\mathrm{Cov}(\sigma_{z}^{(1)},\sigma_{z}^{(2)})\right].\nonumber
\end{eqnarray}
\end{small}
From Eq.~\eqref{FMr}, one can see, $F\sim R^4$ in the case of $R\gg1$, and $F\sim R^2$ in the case of $R\ll1$. In experiment~\cite{Szmuk2015,Lacroute2010}, ${\rho}\sim10^{-8}$ $\mathrm{m}$, thus, $F\sim R^4$ under the condition of $r\gg10^{-8}$ $\mathrm{m}$, $F\sim R^2$ under the condition of $r\ll10^{-8}$ $\mathrm{m}$. Besides, from the expression of QFI, we also find that, the QFI is independent of parameter $\Omega$ to be estimated.

\section{QFI for different initial states}
 From the general expression of QFI in Eq.~\eqref{FMM}, we consider increasing the QFI by searching appropriate states to achieve larger $\gamma$. In this section, we consider two initial states which guarantee $\gamma>0$ to enhance the rotation sensitivity.
\subsection{Partially entangled states }
As a start, we first consider the following partially entangled state
\begin{equation}\label{eq:a,a}
\left|\psi\right\rangle_{\alpha,\alpha}^{\left(n\right)}=\frac{1}{\sqrt{2}}\left[\left(D(\alpha)\left|\uparrow,n\right\rangle\right)^{\bigotimes N}+\left(D(\alpha)\left|\downarrow,n\right\rangle\right)^{\bigotimes N}\right],
\end{equation}
where $D(\alpha)$ is the displacement operator, $\left|n\right\rangle$ is Fock state. For this state, the internal states of each particle are entangled, while each external bosonic state is the displaced Fock state $D(\alpha)\left|n\right\rangle$.
For $\alpha=n=0$, the state reduces to the one considered in Ref.~\cite{Luo2017}.  State $\left|\psi\right\rangle_{\alpha,\alpha}^{\left(n\right)}$ is partially entangled because all the spins are entangled together and all the bosonic modes are not entangled with each other, as well as not with spins.

For state $\left|\psi\right\rangle_{\alpha,\alpha}^{\left(n\right)}$, the corresponding correlation functions are given in Tab.~\ref{tab:tab1}. From the Tab.~\ref{tab:tab1}, one can see that the corresponding QFI is of dependence on the particle number $N$ and $N^2$. Meanwhile, it is of square and biquadratic dependence on the radius $r$.

Based on Eq.~\eqref{FMM}, the corresponding  QFI  can be obtained by
\begin{equation}\label{eq:QFIaa}
F^{\left(n\right)}_{\alpha,\alpha}=4(2n+1)NT_{C}^{2}|C_{1}(\tau)|^2+4N^{2}T_{S}^2 C_{2}^{2}(\tau).
\end{equation}
Additionally, for $n=0$, the state $\left|\psi\right\rangle_{\alpha,\alpha}^{\left(n\right)}$ in Eq.~\eqref{eq:a,a} becomes
\begin{eqnarray}
\left|\psi\right\rangle_{\alpha,\alpha}&=&\frac{1}{\sqrt{2}}\left[\left|\uparrow,\alpha\right\rangle^{\bigotimes N}+\left|\downarrow,\alpha\right\rangle^{\bigotimes N}\right].
\end{eqnarray}
The corresponding QFI expressed in Eq.~\eqref{eq:QFIaa} reduces to
\begin{equation}\label{eq:Faa}
F_{\alpha,\alpha}=4NT_{C}^{2}|C_{1}(\tau)|^2+4N^2T_{S}^2 C_{2}^{2}(\tau).
\end{equation}

From the above results, the QFI for the initial state $\left|\psi\right\rangle_{\alpha,\alpha}^{\left(n\right)}$ is related to the quanta $n$ in each trap mode, but it has no relation with $\alpha$. In other words, the displacement operator $D(\alpha)$ has no influence on the QFI,
i.e., for arbitrary $\alpha$,
\begin{equation}\label{eq:Fn1}
F_{\alpha,\alpha}^{(n)}=F_{0,0}^{(n)}.
\end{equation}
The proof is given as following.
Because of the relations
\begin{eqnarray}
D^{\dagger}(\alpha)aD(\alpha)=a+\alpha, \\
D^{\dagger}(\alpha)a^{\dagger}D(\alpha)=a^{\dagger}+\alpha^{*},
\end{eqnarray}
we have
\begin{eqnarray}\label{eq:hm}
{\cal{H}}^{'}_{M}&=&\left[{D\left(\alpha\right)^{\dagger}}^{\bigotimes N}\right]{\cal{H}}_{M}\left[D(\alpha)^{\bigotimes N}\right] \nonumber \\
&=&{\cal{H}}_{M}+ N T_{c}(C_1(\tau)\alpha^{*}+h.c.).
\end{eqnarray}
Obviously, the second term in Eq.~\eqref{eq:hm} is a constant.
Thus, for the state in Eq.~\eqref{eq:a,a}, one have
\begin{equation}
\bigtriangleup^2{{\cal{H}}^{'}_{M}}=\bigtriangleup^2{\cal{H}}_{M}.
\end{equation}
And based on Eq.~\eqref{eq:QFI}, Eq.~\eqref{eq:Fn1} follows.
Furthermore, if the Fock state $|n\rangle$ in Eq.~\eqref{eq:a,a} is replaced by an arbitrary initial state $|\psi'\rangle$, the independence of $\alpha$ still holds.

Up to now, we have studied the QFI for state $\left|\psi\right\rangle_{\alpha,\alpha}^{\left(n\right)}$.
From Tab.~\ref{tab:tab1}, we know that, for this state, only one spin-spin correlation function in $\gamma$ exists. However, if all three correlation functions in $\gamma$ exist, the QFI may be increased effectively. Thus, in next subsection,  we consider to use a globally entangled state as the initial input.

\begin{table}[h]
\footnotesize
\begin{tabular}{ccc}
\hline
\hline
 & $\left|\psi\right\rangle _{\alpha,\alpha}^{\left(n\right)}$ & $\left|\psi\right\rangle _{\alpha,-\alpha}$\tabularnewline
\hline
$\mathrm{Var}\left(X_{1}\right)$ & $\left(2n+1\right)\left|C_{1}(\tau)\right|^{2}$& $4\left[\textrm{Re}\left(C_{1}(\tau)\alpha^{*}\right)\right]^2+\left|C_{1}(\tau)\right|^{2}$\tabularnewline
$\mathrm{Var}(\sigma_{z}^{\left(1\right)})$ & 1 & 1\tabularnewline
$\mathrm{Cov}(X_{1},\sigma_{z}^{\left(1\right)})$ & 0 & $2\textrm{Re}\left(C_{1}(\tau)\alpha^{*}\right)$\tabularnewline
$\mathrm{Cov}\left(X_{1},X_{2}\right)$ & 0 & $4\left[\textrm{Re}\left(C_{1}(\tau)\alpha^{*}\right)\right]^2$\tabularnewline
$\mathrm{Cov}(\sigma_{z}^{\left(1\right)},\sigma_{z}^{\left(2\right)})$ & 1 & 1\tabularnewline
$\mathrm{Cov}(X_{1},\sigma_{z}^{\left(2\right)})$ & 0 & $2\textrm{Re}\left(C_{1}(\tau)\alpha^{*}\right)$\tabularnewline
\hline
\hline
\end{tabular}
\caption{Correlation functions for initial states $\left|\psi\right\rangle _{\alpha,\alpha}^{\left(n\right)}$ and $\left|\psi\right\rangle_{\alpha,-\alpha}$.}\label{tab:tab1}
\end{table}
\subsection{Globally entangled states}
Based on the idea for the proposed state in Eq.~\eqref{psi1}, we consider the following globally entangled multi-particle state
\begin{equation}\label{eq:a,-a}
\left|\psi\right\rangle_{\alpha,-\alpha}=\frac{1}{\sqrt{2}}\left(\bigotimes_{k=1}^{N}\left|\uparrow,\alpha\right\rangle_{k}+\bigotimes_{k=1}^{N}\left|\downarrow,-\alpha\right\rangle_{k}\right).
\end{equation}
This multi-particle state has both spin-spin entanglement and space-space entanglement. This state is a special type of the state given in Eq.~\eqref{psi1}.
According to Eq.~\eqref{eq:calHM} and \eqref{eq:a,-a}, the corresponding correlation functions are given in Tab.~\ref{tab:tab1}. From Tab.~\ref{tab:tab1}, one can see that the QFI for initial state $\left|\psi\right\rangle_{\alpha,-\alpha}$ is of dependence on $N$ and $N^2$.  Meanwhile, the QFI is of square, cube and biquadratic dependence on the radius $r$.
Additionally, because all correlation functions exist, we can increase the corresponding QFI by adjusting these correlation functions.

According to  Tab.~\ref{tab:tab1} and Eq.~\eqref{FMM}, the QFI for initial state $\left|\psi\right\rangle_{\alpha,-\alpha}$ is obtained as
\begin{eqnarray}\label{eq:QFINN}
F_{\alpha,-\alpha}
&=&4N^2 \left[2 T_{C}\textrm{Re}\left(C_{1}(\tau)\alpha^{*}\right)+T_{S}C_{2}(\tau)\right]^2 \nonumber \\
&&+4NT_{C}^{2}|C_{1}(\tau)|^2 \label{eq:QFINN2}.
\end{eqnarray}
From the expression of $F_{\alpha,-\alpha}$, one can see that the QFI is proportional to the square of the total particle number $N$ in the large $N$ limit. According to the QCRB in Eq.~\eqref{eq:QCRB}, it is obviously to know that the ultimate limit of $\Omega$ in multi-atom Sagnac interferometer can reach the Heisenberg limit, i.e.,
\begin{equation}
\delta\Omega\propto\frac{1}{N},
\end{equation}
which is shown  in Fig.~1.
From this figure,  we find that the slopes of three lines are approximately equal to 2, which indicates $F_{\alpha,-\alpha}\propto N^2$.

\begin{figure}
{\includegraphics[width=80mm]{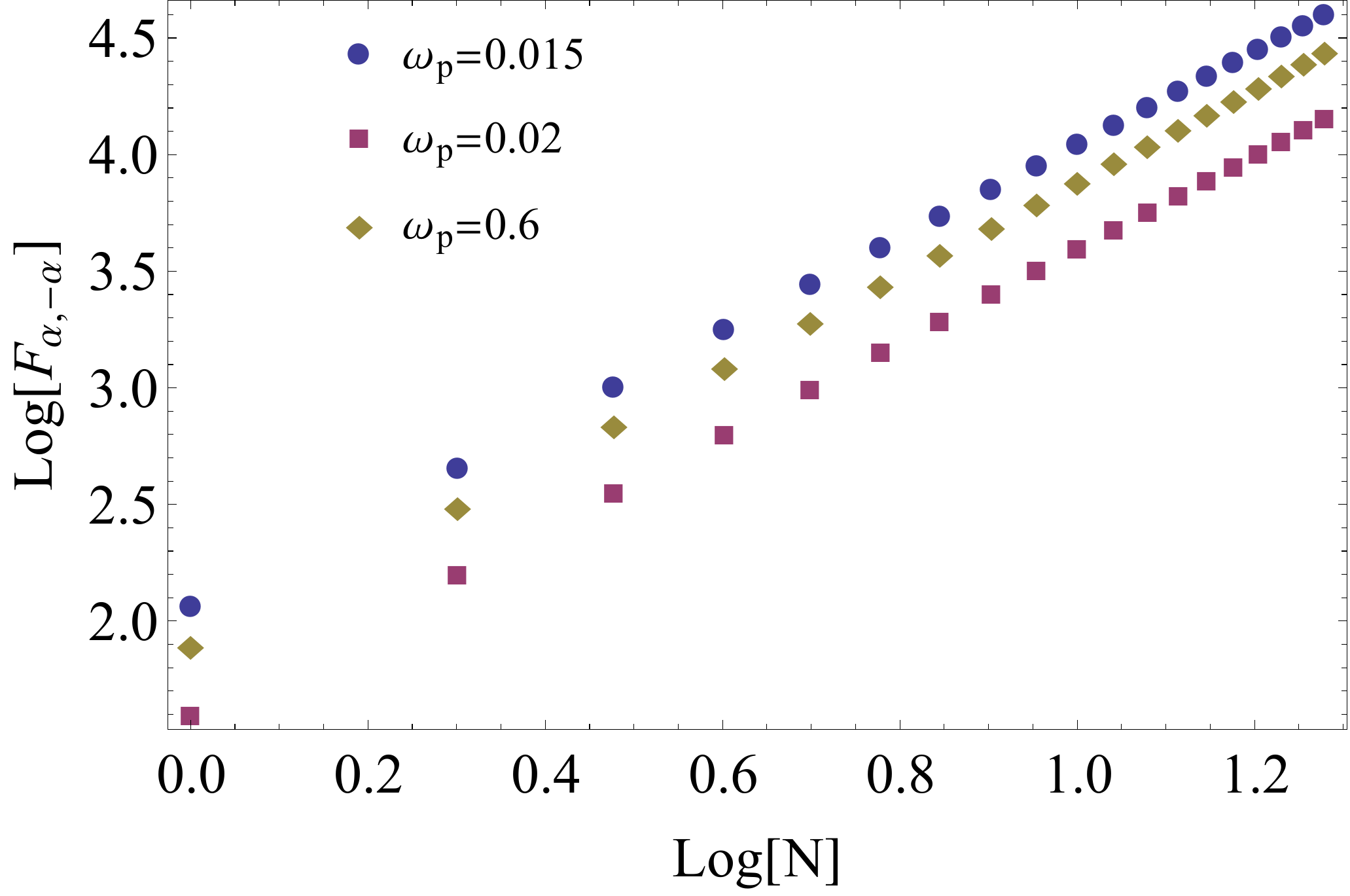}}
\caption{(Color online) The Log-Log plot of the QFI for initial state $\left|\psi\right\rangle_{\alpha,-\alpha}$ versus the total particle number N under different $\omega_{p}$. Here we set $\omega=1$, $m=1$, $\hbar=1$, $r=1$, $\alpha=\mathrm{e}^{i \pi}$. The slopes of the lines are approximately equal to 2, which indicates $F_{\alpha,-\alpha}\propto N^2$.    \label{bbbb}}
\end{figure}
Additionally, from the expression of $F_{\alpha,-\alpha}$, we also find that the QFI for the initial state $\left|\psi\right\rangle_{\alpha,-\alpha}$ depends on the coherent parameter $|\alpha|$ or $\theta_{\alpha}$, where $|\alpha|$ and $\theta_{\alpha}$ are module and argument, respectively. In Fig.~2, the value of QFI varies periodically with $\theta_{\alpha}$. Furthermore, The QFI arrive at the second largest value when $\theta_{\alpha}$ is near to $2z \pi$, where $z$ is integer.  And the QFI gets the maximum value when $\theta_{\alpha}$ is near to $(2z+1)\pi$. Moreover, the QFI increases with the value of $|\alpha|$ under the condition $\theta_{\alpha}=z\pi$.
\begin{figure}
{\includegraphics[width=85mm]{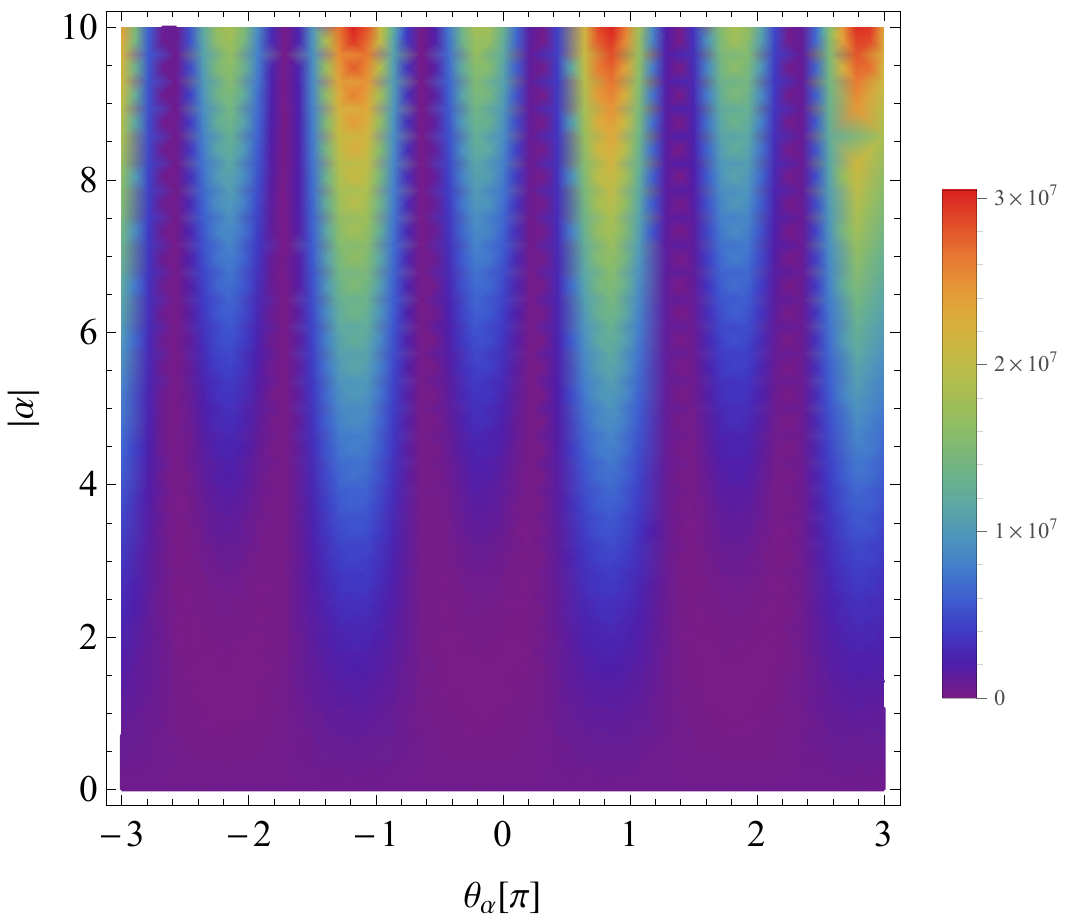}}
\caption{(Color online) QFI as a function of the argument $\theta_{\alpha}$ and the module $|\alpha|$ of the coherent parameter $\alpha$. The unit of $\theta_{\alpha}$ is $\pi$. We set $\omega=1$, $\omega_{p}=0.015 $, $ \tau={\pi}/{\omega_{p}}$, $m=1$, $\hbar=1$, $r=1$, $N=100$. \label{ccc}}
\end{figure}
\subsection{Comparison of the QFI between initial states $\left|\psi\right\rangle_{\alpha,\alpha}$ and $\left|\psi\right\rangle_{\alpha,-\alpha}$.}
In the previous subsections, we have studied the QFI for partially and globally entangled states. In this subsection, we compare the QFI for two initial states to find the one with more advantages in enhancing rotation sensitivity.

By comparing the QFI for states $\left|\psi\right\rangle_{\alpha,\alpha}$ and $\left|\psi\right\rangle_{\alpha,-\alpha}$, we have
\begin{small}
\begin{eqnarray}\label{eq:F1F2}
F_{\alpha,-\alpha}-F_{\alpha,\alpha}&=&16N^2\left[T_{C}\textrm{Re}(C_{1}(\tau){\alpha}^{*})+ T_{S} C_{2}(\tau)\right] \nonumber \\
&&\times\left[T_{C}\textrm{Re}(C_{1}(\tau){\alpha}^{*})\right].
\end{eqnarray}
\end{small}
Because $T_{C}$, $T_{S}$ and $C_{2}(\tau)$ are non-negative real numbers,
in the case of $\textrm{Re}(C_{1}(\tau){\alpha}^{*})\geq0$ or $\textrm{Re}(C_{1}(\tau){\alpha}^{*})\leq-T_{S}C_{2}(\tau)/T_{C}$, the Eq.~\eqref{eq:F1F2} satisfies
\begin{equation}\label{F1-F2}
F_{\alpha,-\alpha}-F_{\alpha,\alpha}\geq0.
\end{equation}
It indicates that the initial state $\left|\psi\right\rangle_{\alpha,-\alpha}$ is more sensitive than state $\left|\psi\right\rangle_{\alpha,\alpha}$ for estimating $\Omega$ under these cases.
For the case of $\textrm{Re}(C_{1}(\tau){\alpha}^{*})\geq0$, if $\alpha$ is a real negative number,
$\textrm{Re}(C_{1}(\tau){\alpha}^{*})=-\alpha\sin^2{\left(\omega\tau/2\right)}\geq0$, then Eq.~\eqref{F1-F2} automatically holds.
For this situation, the globally entangled state is always superior to the locally entangled state in estimating the rotation frequency.

We further numerically study the the difference between
$F_{\alpha,-\alpha}$ and $F_{\alpha,\alpha}$ in Fig.~3.  In Fig.3-(a), $F_{\alpha,-\alpha}/{N^2}$ (red dashed line) and $F_{\alpha,\alpha}/{N^2}$ (blue solid line) are oscillating functions, when $\tau$ takes smaller value. With the increase of $\tau$, $F_{\alpha,\alpha}/{N^2}$ becomes steady, while $F_{\alpha,-\alpha}/{N^2}$ turns into a periodic oscillation with period $T_{0}={2\pi}/{\omega}$. When $\tau=(2l+1)/2 T_{0} , (l=1,2,3...)$, $F_{\alpha,-\alpha}/{N^2}$ reaches the maximum value, the blue line is lower bound of $F_{\alpha,-\alpha}/{N^2}$. The maximum value of $F_{\alpha,-\alpha}/{N^2}$ is approximately three times as high as that of $F_{\alpha,\alpha}/{N^2}$.
\begin{figure}
\subfigure[]{\includegraphics[width=80mm]{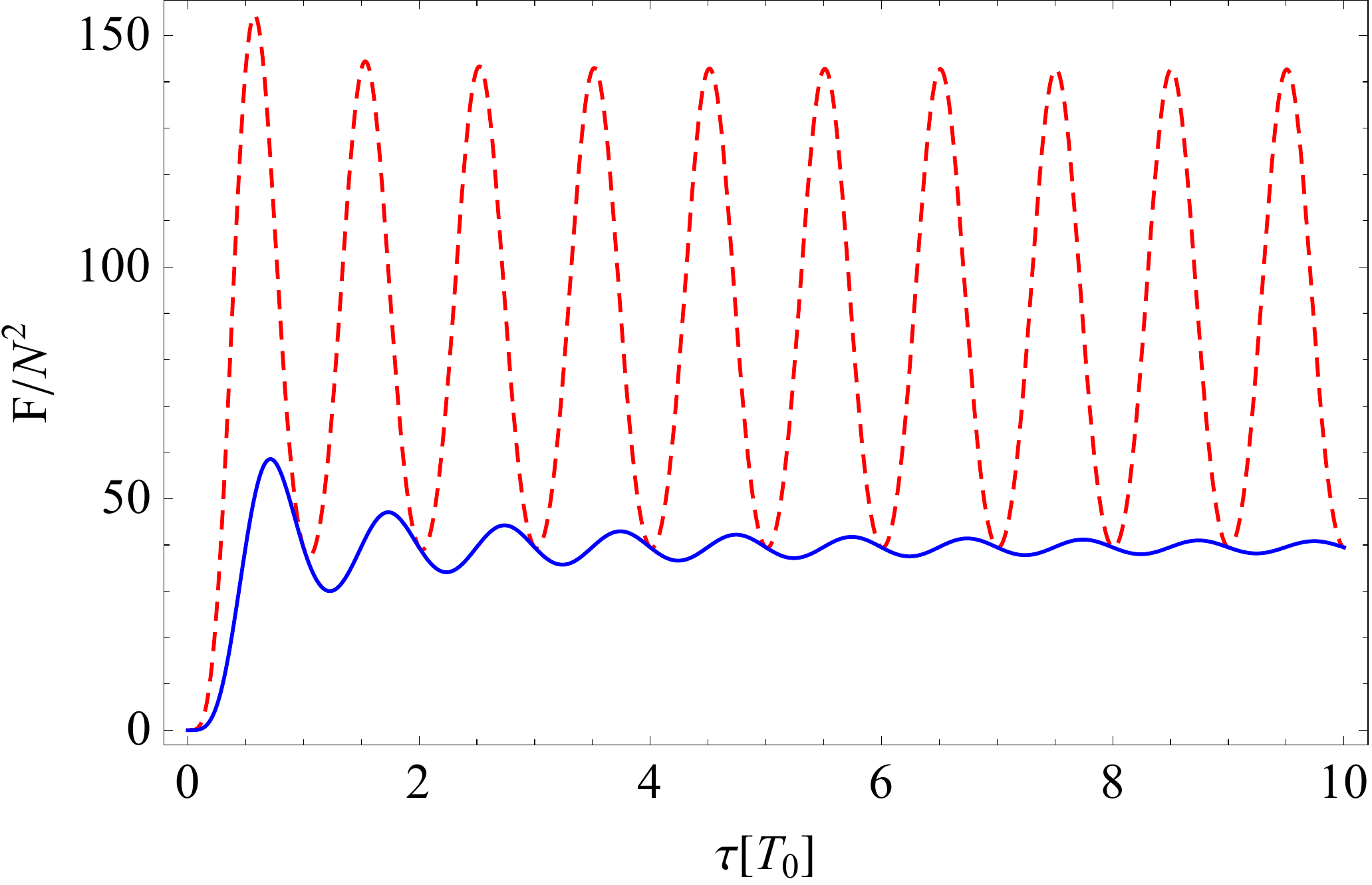}}
\subfigure[]{\includegraphics[width=80mm]{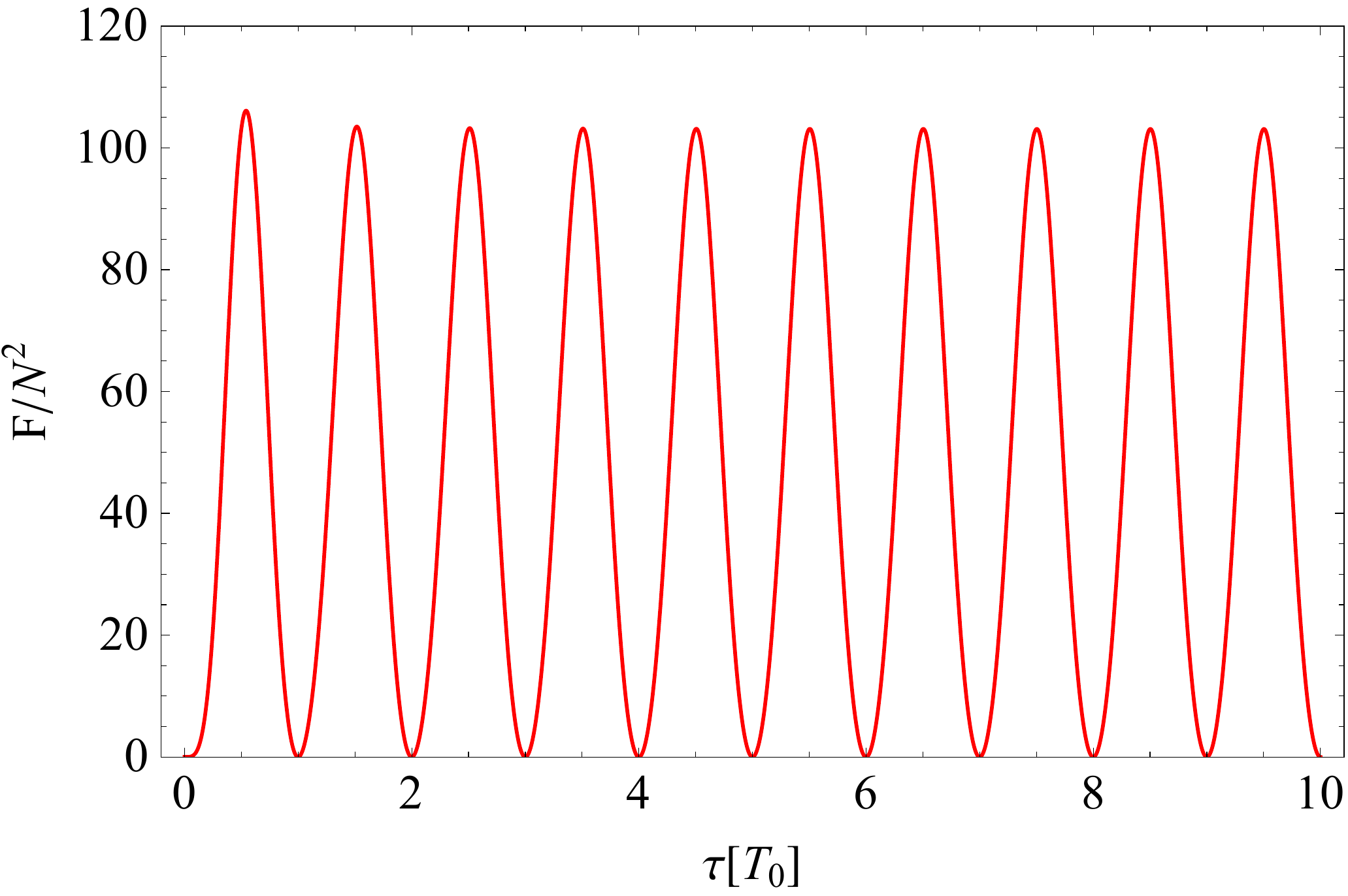}}
\caption{(Color online) (a) $F_{\alpha,-\alpha}/N^2$ (red dashed line) and $F_{\alpha,\alpha}/N^2$ (blue solid line) versus the total evolution time $\tau$. (b) $(F_{\alpha,-\alpha}-F_{\alpha,-\alpha})/N^2$ versus the total evolution time $\tau$. We set $\omega=1$, $\omega_{p}={\pi}/{\tau}$, $m=1$, $\hbar=1$, $r=1$, $\alpha=-1$, $N=100$. The unit of $\tau$ is $T_{0}={2\pi}/{\omega}$.  \label{bbbb}}
\end{figure}
In Fig.3-(b), we give the difference between $F_{\alpha,-\alpha}/{N^2}$ and $F_{\alpha,\alpha}/{N^2}$. $(F_{\alpha,-\alpha}-F_{\alpha,\alpha})/{N^2}$ reach maximum value at $\tau= (2l+1)/2 T_{0}$, while $F_{\alpha,-\alpha}/{N^2}=F_{\alpha,\alpha}/{N^2}$ in the case of $\tau=l T_{0}$.

In addition, in experiment~\cite{Fernholz2007,Kazakov2015,Sherlock2011,Lesanovsky2006}, $\omega_{p}$ is a constant satisfying $\omega_{p}=\pi/\tau$. Based on Eqs.~\eqref{C1} and \eqref{C2}, in the case of $\tau=l T_{0}$, $C_{1}(\tau)=0$, $C_{2}(\tau)=1/2$. Therefore, under this case, $F^{(n)}_{\alpha,\alpha}$, $F_{\alpha,\alpha}$ and $F_{\alpha,-\alpha}$ satisfy the equation
\begin{eqnarray}\label{CQ}
F^{(n)}_{\alpha,\alpha}=F_{\alpha,\alpha}=F_{\alpha,-\alpha}=4N^2 T_{S}^2 C_{2}^2(\tau)\nonumber \\
=4N^{2}\frac{m^{2}{\pi}^2 r^{4}}{\hbar^{2}}=N^{2}\left(\frac{\partial\phi_{S}}{\partial {\Omega}}\right)^2,
\end{eqnarray}
based on Eqs.~\eqref{eq:QFIaa}, \eqref{eq:Faa}, and \eqref{eq:QFINN2}.
It means that the QFI for states $\left|\psi\right\rangle^{(n)}_{\alpha,\alpha}$ and $\left|\psi\right\rangle_{\alpha,-\alpha}$ are equal in the case of $\tau={2\pi l}/{\omega} $.  The reason is that the external part $\exp(-\mathrm{i}\omega {a}^{\dagger} {a}\tau)$ of the evolution operator in Eq.~\eqref{eq:Ut} is a unit matrix  at $\tau={2\pi l}/{\omega} $ for all the three initial state.

\section{Conclusion}
We have studied the multi-particle Sagnac atom interferometer and given an effective way to enhance the sensitivity
of the rotation frequency. An efficient way to study QFI for dynamical processes is via the Hermitian generator method. We have explicitly presented
the generator with respect to the rotation frequency for arbitrary time dependence of the angular speed. This generator contains the well-known Sagnac phase. Taking advantage of the generator, it is very convenient to find the optimal initial state to achieve high precision in estimating the
rotation frequency.

We have derived the general expression of QFI with respect to $\Omega$ for any initial pure states in terms of correlation functions. The QFI was found to be a linear superposition of particle number $N$ and $N^2$, and it is of square, cubic, and biquadratic
dependence on the radius $r$ of the ring. We can enhance the rotation sensitivity by searching appropriate states which guarantees that $N^2$ term is dominant. We proposed to use partially and globally entangled states, respectively. By analysing and comparing the QFI for two initial states, we found that the globally entangled state can be more sensitive than partially entangled state in estimating the rotation frequency. Moreover, the generator obtained in this work is applicable to the case with initial mixed states and the situation of decoherence.

\section{Acknowledgments}
This work was supported by the National Key Research and
Development Program of China (No.~2017YFA0304202 and No.~2017YFA0205700),
the NSFC through Grant No.~11475146, and the Fundamental Research Funds for
the Central Universities through Grant No.~2017FZA3005.

\appendix

\section{The derivation of $\cal{H}_{M}$}
For the Hamiltonian
\begin{equation}
H = H_{0}+H',
\end{equation}
where
\begin{align}
H_{0} & =\hbar\omega a^{\dagger}a,\\
H' & =g\left(t\right)a+g^{\ast}\left(t\right)a^{\dagger}.
\end{align}
In the interaction picture, the time-dependent Hamiltonian $H'$  can be expressed as
\begin{equation}
H_{I}' = g\left(t\right)a\mathrm{e}^{-\mathrm{i}\omega t}+g^{\ast}\left(t\right)a^{\dagger}\mathrm{e}^{\mathrm{i}\omega t}.
\end{equation}
Assume that the evolution operator is
\begin{equation}
U(t)={\mathrm{e}}^{-{\mathrm{i}}\frac{H_{0}}{\hbar}t}U_{I},
\end{equation}
where
\begin{equation}
U_{I}=\mathrm{e}^{\mathrm{i}\Phi\left(t\right)}\mathrm{e}^{\eta\left(t\right)a^{\dagger}-\eta^{*}\left(t\right)a}=\mathrm{e}^{\mathrm{i}\Phi\left(t\right)}D[\eta{(t)}].\label{eq:UI}
\end{equation}
The displacement operator $ D\left[\eta(t)\right]$ satisfies the following equation
\begin{equation}
\partial_{x}D\left[\eta(t)\right]=\left[\left(\partial_{x}{\eta}(t)a^{\dagger}+\frac{1}{2}\eta(t)\partial_{x}{\eta}^{*}(t)\right)-h.c.\right]D[\eta(t)]. \label{eq:DD}
\end{equation}
According to
\begin{equation}
\frac{\partial U_{I}}{\partial t}=-\mathrm{\frac{i}{\hbar}}H_{I}'U_{I},
\end{equation}
we can derive the equation as following,
\begin{eqnarray}
\eta\left(t\right)&=&-\mathrm{\frac{i}{\hbar}}\int_{0}^{t}g^{\ast}\left(s\right)\mathrm{e}^{\mathrm{i}\omega s}ds\\
\Phi\left(t\right)
&=&\frac{\mathrm{i}}{2\hbar^{2}}\int_{0}^{t}dt_{1}\int_{0}^{t_{1}}dt_{2}\big[g^{\ast}\left(t_{2}\right)g\left(t_{1}\right)\mathrm{e}^{\mathrm{i}\omega\left(t_{2}-t_{1}\right)}\nonumber\\
& & -g\left(t_{2}\right)g^{\ast}\left(t_{1}\right)\mathrm{e}^{\mathrm{i}\omega\left(t_{1}-t_{2}\right)}\big].\label{eq:tt}\\ \nonumber
\end{eqnarray}
Therefore, the evolution operator for Hamiltonian $H$ can be obtained by
\begin{eqnarray}
U\left(t\right)=\mathrm{e}^{-\mathrm{i}\omega a^{\dagger}at}\mathrm{e}^{\mathrm{i}\Phi\left(t\right)}D[\eta(t)].\label{eq:Ut}\\ \nonumber
\end{eqnarray}
For the Hamiltonian in Eq.~\eqref{eq:H2}, the evolution operator can be given by
\begin{equation}\label{eq:HH}
U(\tau)=\exp(-\mathrm{i}\omega {a}^{\dagger} {a}\tau)\exp[\mathrm{i}\Phi\left(\sigma_{z},\tau\right)]D[\eta\left(\sigma_{z},\tau\right)],
\end{equation}
where $\tau$ is the total evolution time which satisfying $\int_{0}^{\tau}\omega_{p}(t)dt=\pi$, and
\begin{equation}
\Phi\left(\sigma_{z},\tau\right) = \int_{0}^{\tau}\int_{0}^{t_{1}}f(\sigma_{z},t_{1})f(\sigma_{z},t_{2})\sin[\omega(t_{1}-t_{2})]dt_{2}dt_{1},\label{eq:phi}
\end{equation}
\begin{eqnarray}
\eta\left(\sigma_{z},\tau\right)&=& -\int_{0}^{\tau}f(\sigma_{z},t)\mathrm{e}^{\mathrm{i}\omega t}dt,\label{eq:eta} \\
f(\sigma_{z},t)& = &\sqrt{\frac{m\omega}{2\hbar}}r\big[\Omega+{\sigma}_{z}\omega_{p}(t)\big].\label{eq:f}
\end{eqnarray}
\begin{widetext}
According to Eqs.~\eqref{eq:generator} and \eqref{eq:HH},
the generator in single-atom Sagnac interferometer can be given by
\begin{eqnarray}
\mathcal{H} & = & \mathrm{i}\big\{\partial_{\Omega}D^\dagger\left[\eta\left(\sigma_{z},\tau\right)\right]\big\}D\left[\eta\left(\sigma_{z},\tau\right)\right]
  +\mathrm{i}\big\{\partial_{\Omega}\exp[-\mathrm{i}\Phi\left(\sigma_{z},\tau\right)]\big\}\exp[\mathrm{i}\Phi\left(\sigma_{z},\tau\right)]\nonumber\\
  &=&\mathrm{i}\left\{\left[\partial_{\Omega}\eta^{*}(\sigma_{z},\tau)\right]a-\left[\partial_{\Omega}\eta(\sigma_{z},\tau)\right]a^{\dagger}
  +\frac{1}{2}\big[\eta^{*}(\sigma_{z},\tau)\partial_{\Omega}\eta(\sigma_{z},\tau)-\eta(\sigma_{z},\tau)\partial_{\Omega}\eta^{*}(\sigma_{z},\tau)
\big]\right\}+\partial_{\Omega}\Phi\left(\sigma_{z},\tau\right).\label{eq:a} \\ \nonumber
  \end{eqnarray}
From the Eqs.~\eqref{eq:eta} and \eqref{eq:f}, we can successively obtain two equations as following,
\begin{equation}
\partial_{\Omega}\eta(\sigma_{z},\tau) =-r\sqrt{\frac{2m}{\hbar\omega}}\sin\left(\frac{\omega \tau}{2}\right)\exp(\frac{\mathrm{i}\omega \tau}{2}),
\end{equation}
\begin{equation}
\mathrm{i}\left\{\frac{1}{2}\eta^{*}(\sigma_{z},\tau)\left[\partial_{\Omega}\eta(\sigma_{z},\tau)\right]-\frac{1}{2}\eta(\sigma_{z},\tau)\left[\partial_{\Omega}\eta^{*}(\sigma_{z},\tau)\right]
\right\}
=\frac{\sigma_{z}mr^{2}}{2\hbar}\int_{0}^{\tau}\omega_{p}\left(t\right)\big\{\cos\left(\omega t\right)-\cos\left[\omega(t-\tau)\right]\big\}dt.
\end{equation}
Based on Eq.~\eqref{eq:phi}, we obtain
\begin{eqnarray}
\partial_{\Omega}\Phi\left(\sigma_{z},\tau\right)
& = & \frac{m\omega}{2\hbar}r^{2}\int_{0}^{\tau}\int_{0}^{t_{1}}\big\{2\Omega\sin\left[\omega\left(t_{1}-t_{2}\right)\right]
 +\sigma_{z}\omega_{p}\left(t_{1}\right)\sin\left[\omega\left(t_{1}-t_{2}\right)\right]
 +\sigma_{z}\omega_{p}\left(t_{2}\right)\sin\left[\omega\left(t_{1}-t_{2}\right)\right]\big\}dt_{2}dt_{1}\nonumber \\
& = & \frac{mr^{2}}{2\hbar}\left\{2\Omega\left(\tau-\frac{\sin\omega \tau}{\omega}\right)+2\sigma_{z}\pi-\sigma_{z}\int_{0}^{\tau}\omega_{p}\left(t_{1}\right)\big[\cos\omega t_{1}
+\cos(\omega\left(\tau-t_{1}\right))dt_{1}\big]\right\}.
\end{eqnarray}
Inserting above three equations into the Eq.~\eqref{eq:a}, the generator of QFI with respect to $\Omega$ in the single-atom Sagnac interferometer can be given by
\begin{equation}
\mathcal{H}=T_{C}\left[C_1(\tau)a^{\dagger}+h.c.\right]+\frac{1}{\omega}C_{0}(\tau)+T_{S}C_2(\tau)\sigma_{z}.\\
\end{equation}

For the Hamiltonian of multi-atom Sagnac interferometer in Eq.~\eqref{eq:HM}, the evolution operator is
\begin{equation}
U_{M}(\tau)=U_{1}(\tau)U_{2}(\tau)...U_{k}(\tau)...U_{N}(\tau).
\end{equation}
The generator of QFI with respect to $\Omega$ in multi-atom Sagnac interferometer can be obtained by
\begin{equation}
{\cal{H}}_{M}=\mathrm{i}\left(\partial_{\Omega}U_{M}^{\dagger}(\tau)\right)U_{M}(\tau)=\sum_{k=1}^{N}{\cal{H}}_{S,k}.
\end{equation}
\section{The derivation of the general QFI with respect to $\Omega$ in multi-atom interferometer}
The covariance of two observable quantities $A,B$ is given by the following formula
\begin{equation}
\mathrm{Cov}(A,B)=\left\langle\frac{AB+BA}{2}\right\rangle-\langle A\rangle \langle B\rangle.
\end{equation}
In above equation, $\displaystyle A=\sum_{k}A_{k}$, $\displaystyle B=\sum_{k}B_{k}$, then we express the covariance explicitly
\begin{equation}
\mathrm{Cov}(A,B)=\mathrm{Cov}\left(\sum_{k} A_{k},\sum_{k'}B_{k'}\right)
=\sum_{kk'}\mathrm{Cov}(A_{k},B_{k'})=\sum_{k}\mathrm{Cov}(A_{k},B_{k})+\sum_{k\neq k'}\mathrm{Cov}(A_{k},B_{k'}).
\end{equation}
Note that the symmetry for above equation implies
\begin{equation}
\mathrm{Cov}(A_{k},B_{k'})=\begin{cases}\mathrm{Cov}(A_{1},B_{1})&k=k'\\ \mathrm{Cov}(A_{1},B_{2})&k\neq k'\end{cases},
\end{equation}
Thus,
\begin{equation}\label{Eq:Cov}
\mathrm{Cov}(A,B)=N\mathrm{Cov}(A_{1},B_{1})+(N^{2}-N)\mathrm{Cov}(A_{1},B_{2}).
\end{equation}
Especially, when $A_{k}=B_{k}$, the above equation become
\begin{equation}\label{Eq:Var}
\mathrm{Var}\left(\sum_{k} A_{k}\right)=N\mathrm{Var}(A_{1})+(N^{2}-N)\mathrm{Cov}(A_{1},A_{2}),
\end{equation}
where $\mathrm{Var}(A_{1})=\left\langle A_{1}^{2}\right\rangle -\left\langle A_{1}\right\rangle ^{2}$ is the variance of $A_{1}$.

Based on the Eqs.~\eqref{Eq:Cov} and \eqref{Eq:Var}, we obtain $\mathbb{\mathrm{Var}}\left({\cal H}_{1}\right)$, $\mathbb{\mathrm{Var}}\left({\cal H}_{2}\right)$ and $\mathbb{\mathrm{Cov}}\left({\cal H}_{1}{\cal H}_{2}\right)$ as following,
\begin{eqnarray}
\mathrm{Var}({\cal{H}}_{1})&=&\mathrm{Var}\left(T_{C}\sum_{k=1}^{N}X_{k}\right)=T_{C}^{2}N\mathrm{Var}(X_{1})+T_{C}^2(N^2-N)\mathrm{Cov}(X_{1},X_{2}),\\
\mathrm{Var}({\cal{H}}_{2})&=&\mathrm{Var}\left(T_{S}C_{2}(\tau)\sum_{k=1}^{N}\sigma_{z}^{(k)}\right)=T_{S}^2 C_{2}^2(\tau)N\mathrm{Var}(\sigma_{z}^{(1)})+T_{S}^2 C_{2}^2(\tau)(N^2-N)\mathrm{Cov}(\sigma_{z}^{(1)},\sigma_{z}^{(2)}),\\
\mathrm{Cov}({\cal{H}}_{1},{\cal{H}}_{2})&=&T_{C}T_{s}C_{2}(\tau)N\mathrm{Cov}(X_{1},\sigma_{z}^{(1)})+T_{C}T_{S}C_{2}(\tau)(N^2-N)\mathrm{Cov}(X_{1},\sigma_{z}^{(2)}).
\end{eqnarray}
According to Eqs.~\eqref{eq:QFI} and \eqref{eq:calHM}, when the initial state is a pure state, the corresponding QFI with respect to $\Omega$ can be given by Eqs.~\eqref{FM} and \eqref{FMM}
\end{widetext}

\end{document}